\documentclass[letter]{aa}  

\usepackage{graphicx}
\usepackage{multirow}
\usepackage{lineno}
\usepackage{xcolor}
\usepackage{soul}
\usepackage{placeins}
\usepackage{txfonts}
\begin{document} 

   \title{Doppler wind measurements in Neptune's stratosphere with ALMA}

   \author{\'Oscar Carri\'on-Gonz\'alez\inst{1}\thanks{\email{oscar.carrion@obspm.fr, oscar.carrion.gonzalez@gmail.com}}
         \and Raphael Moreno\inst{1}
         \and Emmanuel Lellouch\inst{1}
         \and Thibault Cavali\'e\inst{2,1}
         \and Sandrine Guerlet\inst{3,1}
         \and Gwena\"el Milcareck\inst{3}
         \and Aymeric Spiga\inst{3}
         \and No\'e Cl\'ement\inst{2}
         \and J\'er\'emy Leconte\inst{2}
         }

   \institute{
        LESIA, Observatoire de Paris, Universit\'e PSL, CNRS, Sorbonne Universit\'e, Universit\'e Paris Cit\'e, 5 place Jules Janssen, 92195 Meudon, France
         \and
         Laboratoire d’Astrophysique de Bordeaux, Univ. Bordeaux, CNRS, B18N, All\'ee Geoffroy Saint-Hilaire, 33615 Pessac, France
         \and
         Laboratoire de M\'et\'eorologie Dynamique/Institut Pierre-Simon Laplace (LMD/IPSL), Sorbonne Universit\'e, CNRS, \'Ecole Polytechnique, Institut Polytechnique de Paris, \'Ecole Normale Sup\'erieure (ENS), PSL Research University, Paris, France}

 
 \abstract
   {Neptune's tropospheric winds are among the most intense in the Solar System, but the dynamical mechanisms that produce them remain uncertain. Measuring wind speeds at different pressure levels may help understand the atmospheric dynamics of the planet.}
   {The goal of this work is to directly measure winds in Neptune’s stratosphere with ALMA Doppler spectroscopy.} 
   {We derived the Doppler lineshift maps of Neptune at the CO(3-2) and HCN(4-3) lines at 345.8~GHz ($\lambda\sim0.87$~mm) and 354.5~GHz (0.85~mm), respectively. For that, we used spectra obtained with ALMA in 2016 and recorded with a spatial resolution of $\sim$0.37" on Neptune's 2.24" disk. After subtracting the planet solid rotation, we inferred the contribution of zonal winds to the measured Doppler lineshifts at the CO and HCN lines. We developed an MCMC-based retrieval methodology to constrain the latitudinal distribution of wind speeds.}
   {We find that CO(3-2) and HCN(4-3) lines probe the stratosphere of Neptune at pressures of $2^{+12}_{-1.8}$~mbar and $0.4^{+0.5}_{-0.3}$~mbar, respectively.
   The zonal winds at these altitudes are less intense than the tropospheric winds based on cloud tracking from Voyager observations.
   We find equatorial retrograde (westward) winds of $-180^{+70}_{-60}$~m/s for CO, and $-190^{+90}_{-70}$~m/s for HCN.
   Wind intensity decreases towards mid-latitudes, and wind speeds at 40$^\circ$S are $-90^{+50}_{-60}$~m/s for CO, and $-40^{+60}_{-80}$~m/s for HCN.
   Wind speeds become 0~m/s at about 50$^\circ$S, and we find that the circulation reverses to a prograde jet southwards of 60$^\circ$S.
   Overall, our direct stratospheric wind measurements match previous estimates from stellar occultation profiles and expectations based on thermal wind equilibrium.}
   {These are the first direct Doppler wind measurements performed on the Icy Giants, opening a new method to study and monitor their stratospheric dynamics.}


   \maketitle
%

\section{Introduction} \label{sec:introduction}

Voyager 2 measurements of Neptune's atmosphere revealed some of the most intense zonal winds ever measured in the Solar-System planets \citep{hammeletal1989, smithetal1989}.
This was done by tracking cloud motions in images of Voyager's ISS instrument, and constrained Neptune winds to be $-400$~m/s (retrograde) at the equator, with a prograde jet of about +~250~m/s at latitude 70ºS \citep{limaye-sromovsky1991, sromovskyetal1993}.
Subsequent cloud-tracking measurements with the Hubble Space Telescope \citep{sromovskyetal2001a, sromovskyetal2001b} and ground-based observatories with adaptive optics \citep{fitzpatricketal2014, tollefsonetal2018} confirmed this general wind pattern \citep{fletcheretal2020}.

Cloud-tracking methods, however, generally lack an accurate constraint of the atmospheric altitudes probed by the observations.
Although clouds are mostly expected to be located in the upper troposphere ($\sim$100~mbar--1~bar), determining the exact cloud-top pressure levels requires multiple-scattering radiative-transfer computations.
These are highly dependent on the assumed optical properties of the atmospheric aerosols, and on the assumed vertical distribution of gaseous species and aerosols \citep[e.g.][]{luszczcooketal2016}.
Furthermore, cloud pressure levels have been found to vary with latitude by about an order of magnitude \citep{irwinetal2011, irwinetal2016, huesoetal2017}.
Even clouds at similar latitudes have been found to be located at pressure levels as different as 30 and 300~mbar \citep{depateretal2014}.

Above the cloud level, wind information has been derived primarily from the thermal wind equation that relates the latitudinal temperature gradient to the vertical wind shear.
Temperature fields from Voyager/IRIS spectra were initially used for this \citep{conrathetal1989}.
Later, additional ground-based thermal observations became available \citep{fletcheretal2014}. At tropospheric levels, these data consistently indicate a warm equator, cool mid-latitudes and a warmer south pole. Albeit it cannot be applied at the equator, the thermal wind equation implies decaying winds with altitude at low latitudes (-30° to 30°) and essentially altitude-independent winds at mid-to-high southern latitudes.
Relatively small changes were found in the predicted circulation pattern from 1989 to 2003 (southern summer solstice), except near the south pole, found to be anomalously warm (by $\sim$5~K) in 2003 \citep{ortonetal2007}.

Additional wind measurements with other techniques were performed for a reduced number of latitudes.
\citet{frenchetal1998} measured stratospheric winds from three stellar occultations over 1985-1989, constraining the wind-distorted shape of the planet. 
They inferred that wind speeds at the 0.38 mbar level are 0.6$\pm$0.2 times those of Voyager cloud-tracking from \citet{sromovskyetal1993}.
Also, they found tentative evidence of winds at 0.7 $\mu$bar further decaying to $\sim$0.17 times the tropospheric values. This was the first direct evidence for the decay of winds above the tropopause, with an estimated wind shear in agreement with that inferred from Voyager temperature fields at deeper levels.

In this work we aim to directly measure Neptune's stratospheric winds based on Doppler lineshift measurements of CO and HCN lines with the Atacama Large Millimeter/submillimeter Array (ALMA). 
Millimetric and submillimetric measurements have already been used to directly measure the winds in the atmospheres of Venus \citep[e.g.][and references therein]{lellouchetal2008}, Mars \citep{lellouchetal1991, cavalieetal2008, morenoetal2009}, Jupiter \citep{cavalieetal2021}, Saturn \citep{benmahietal2022}, and Titan \citep{morenoetal2005, lellouchetal2019, cordineretal2020}.
No such analysis was available for the Icy Giants to date.
Besides probing the otherwise difficult-to-sound stratosphere, an advantage of the technique is that it is insensitive to aerosols, reducing considerably the uncertainties on the probed levels.

\section{Observations} \label{sec:observations}

\begin{figure*}
  	\centering
	\includegraphics[width=13.25cm]{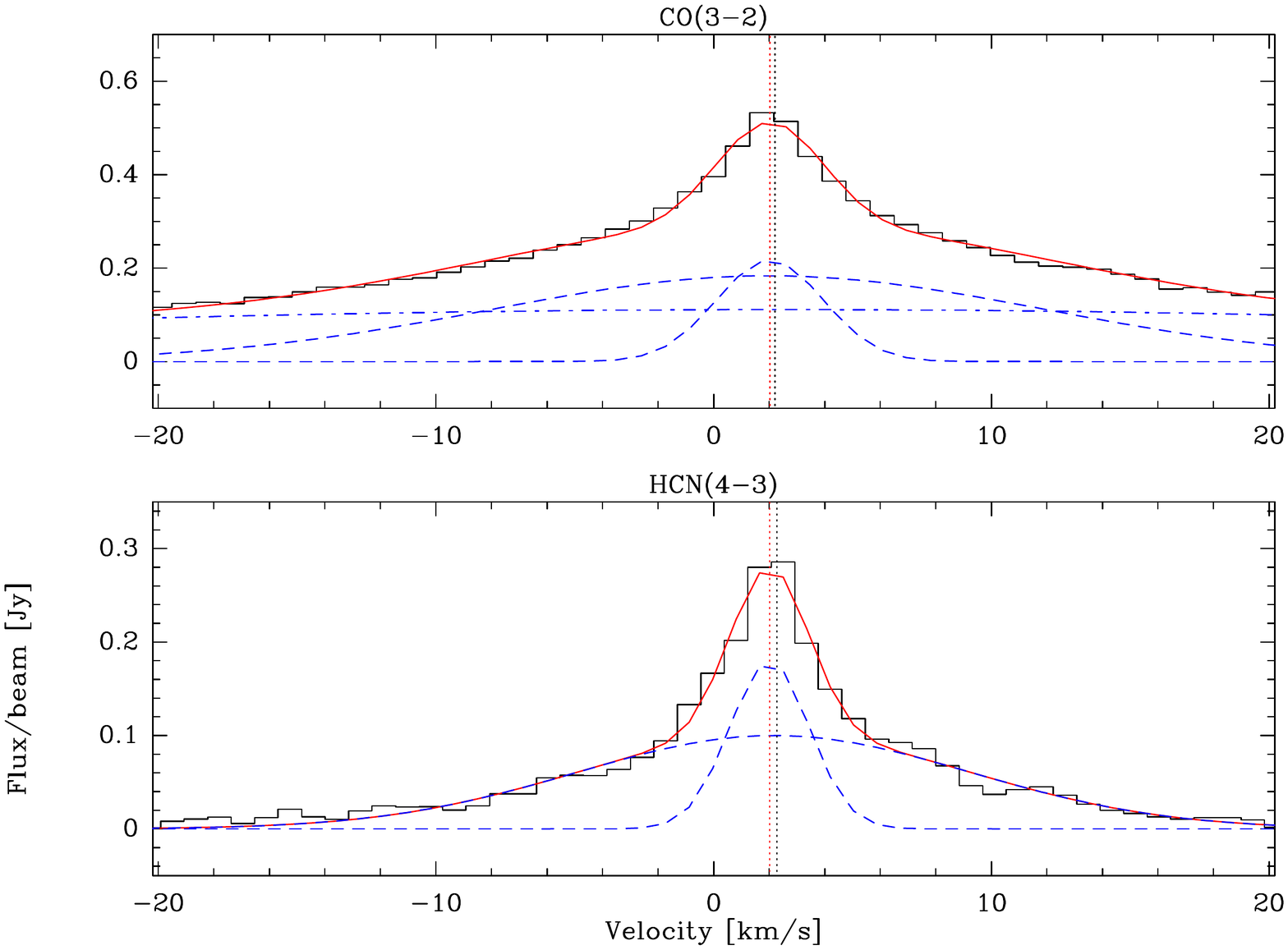} 
    \hspace{-1.25cm}
    \includegraphics[width=6cm]{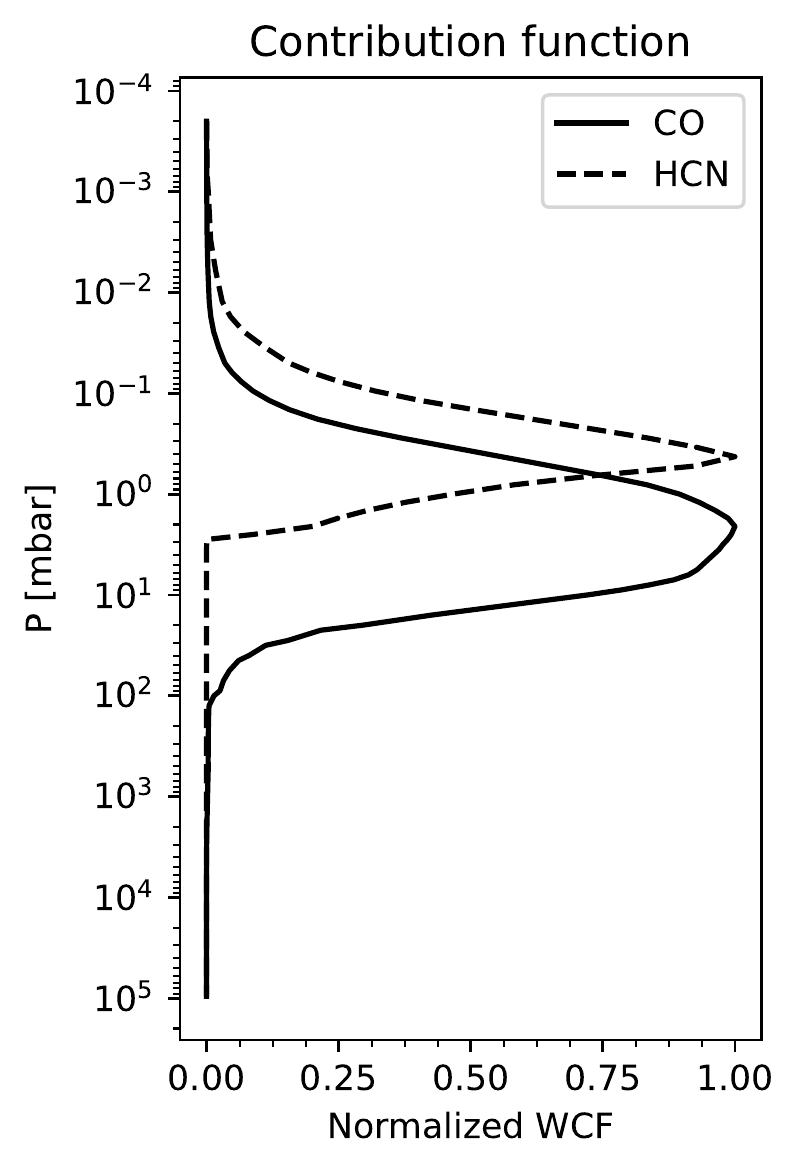} 
	\caption{Example of measured spectra (left column) of CO(3-2) and HCN(4-3) lines (black) at the position of the sky western equatorial limb, at an oﬀset from Neptune centre (-1.1", 0.0"). Individual Gaussian fit components with FWHMs of about 4, 20, 80 km/s are shown in dashed blue, and their sum in red. Also shown, the beam-convolved solid body velocity (black dotted line) and the fit velocity (red dotted line). The difference of these two velocities allows us to derive the Doppler winds.
    Right: Normalized wind contribution functions (WCF) at the limb for each molecule. }
\label{fig:observations}
\end{figure*}

\begin{figure*}
	\centering
	\includegraphics[width=5.95cm]{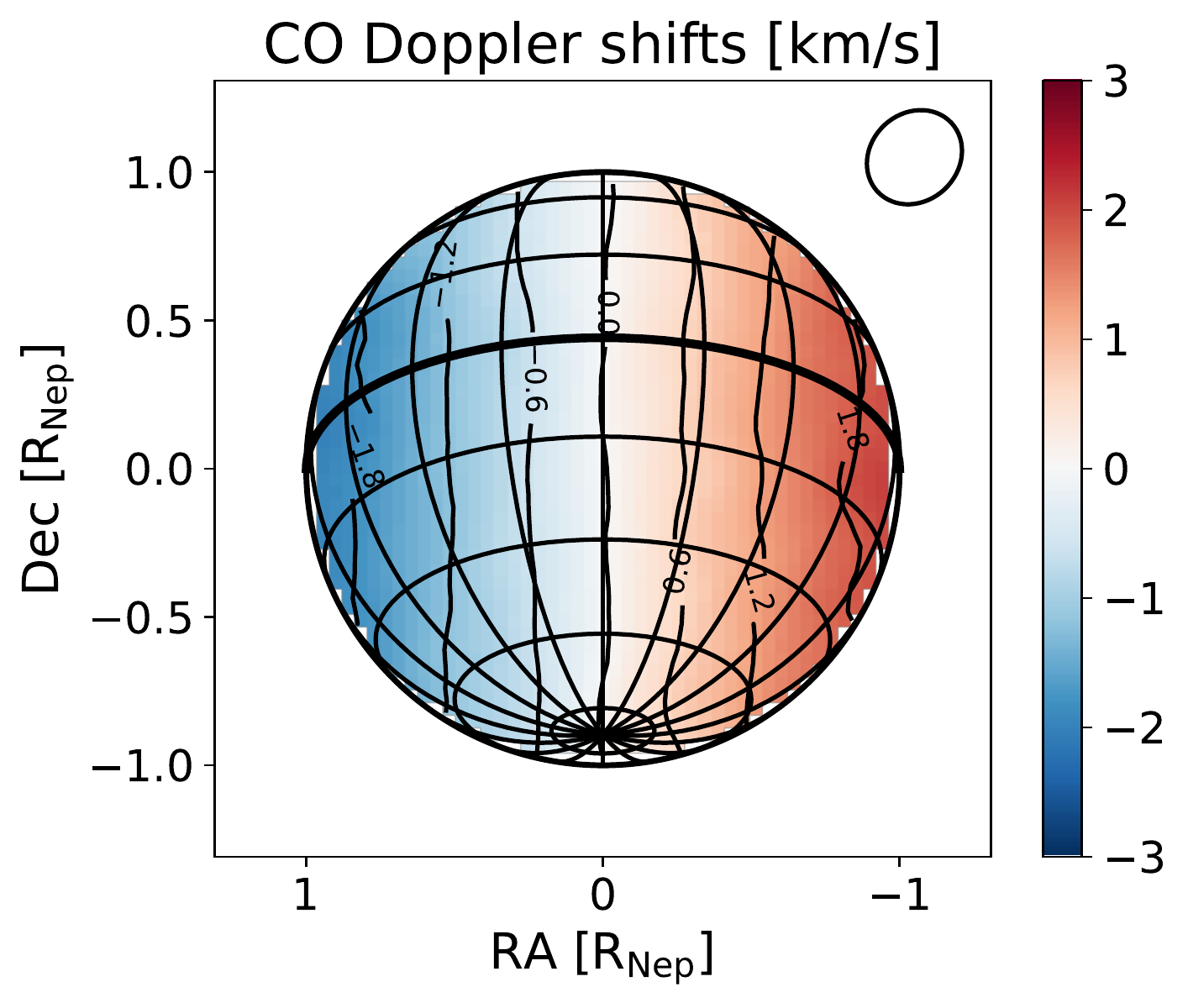} 
	\hfill
	\includegraphics[width=6.125cm]{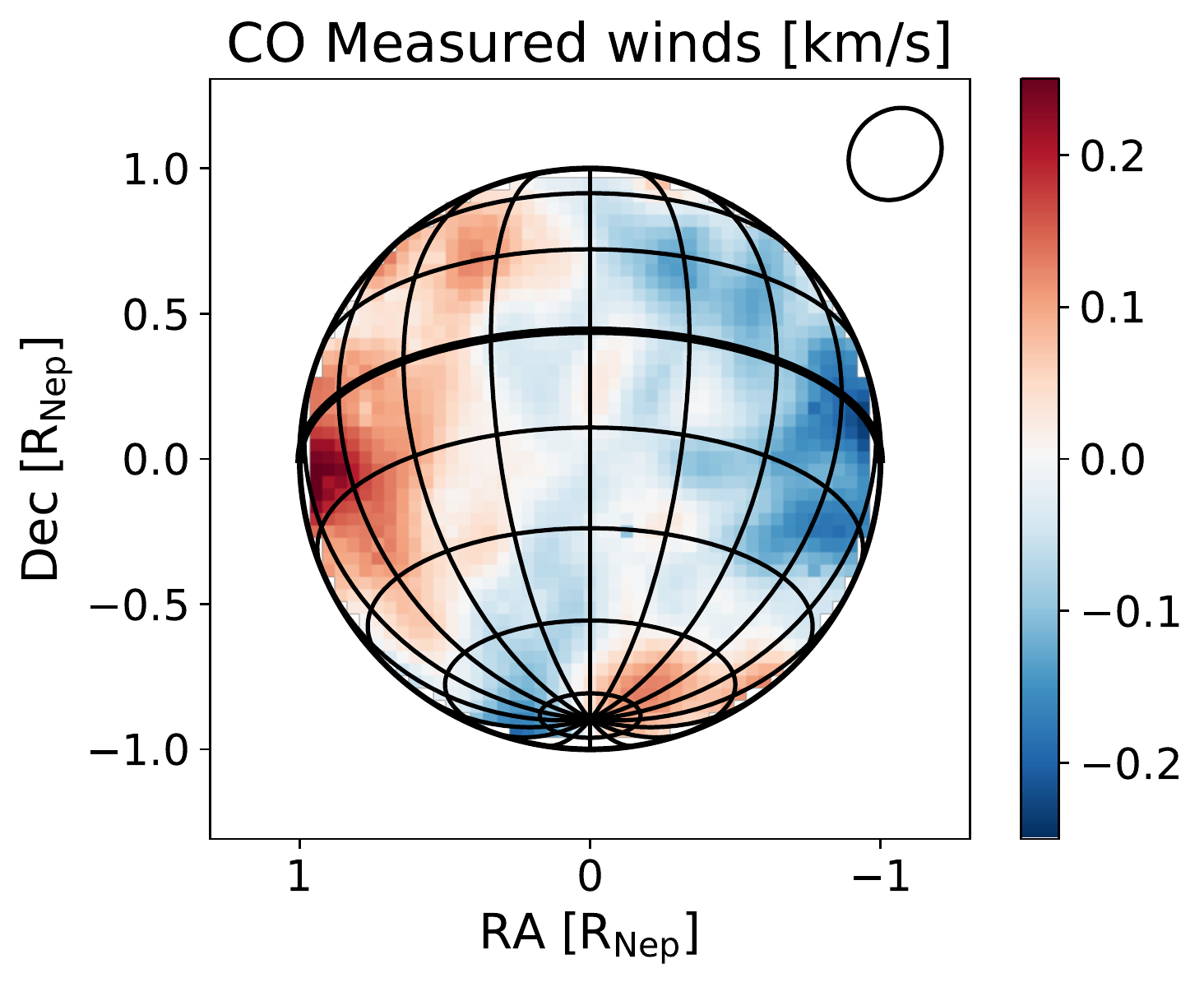} 
	\hfill
	\includegraphics[width=6.125cm]{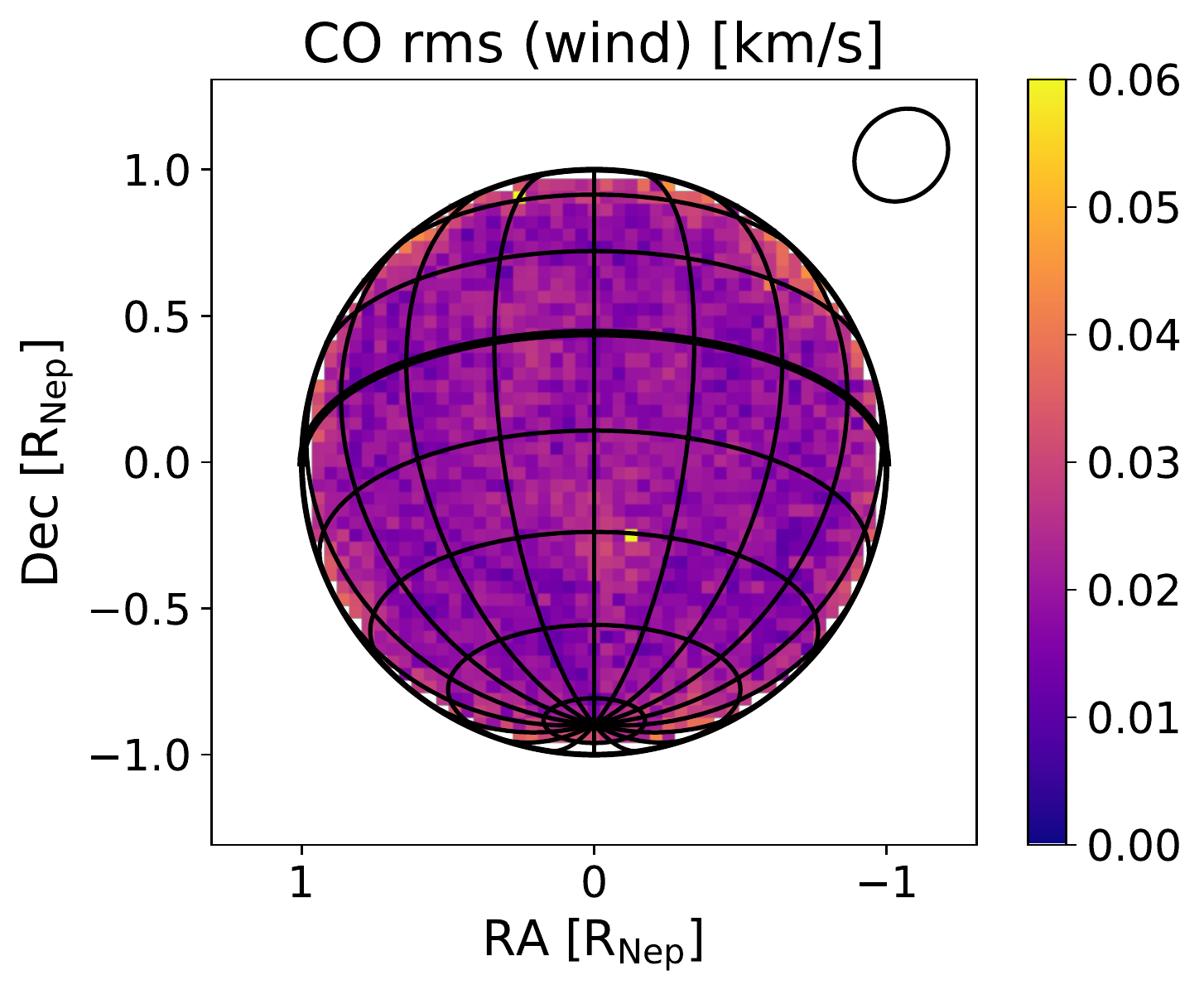} 
	   \\
	\includegraphics[width=5.95cm]{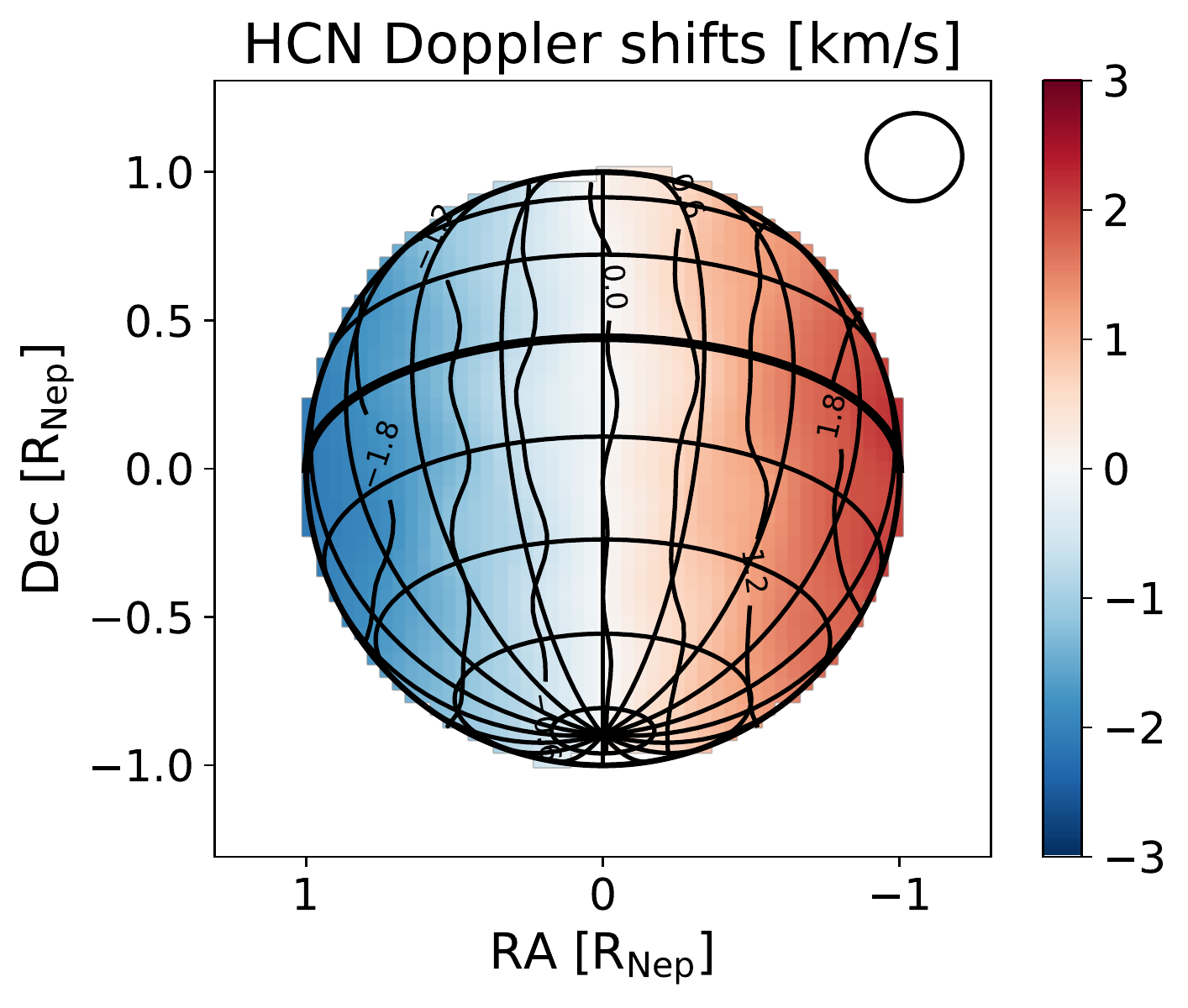} 
	\hfill
	\includegraphics[width=6.125cm]{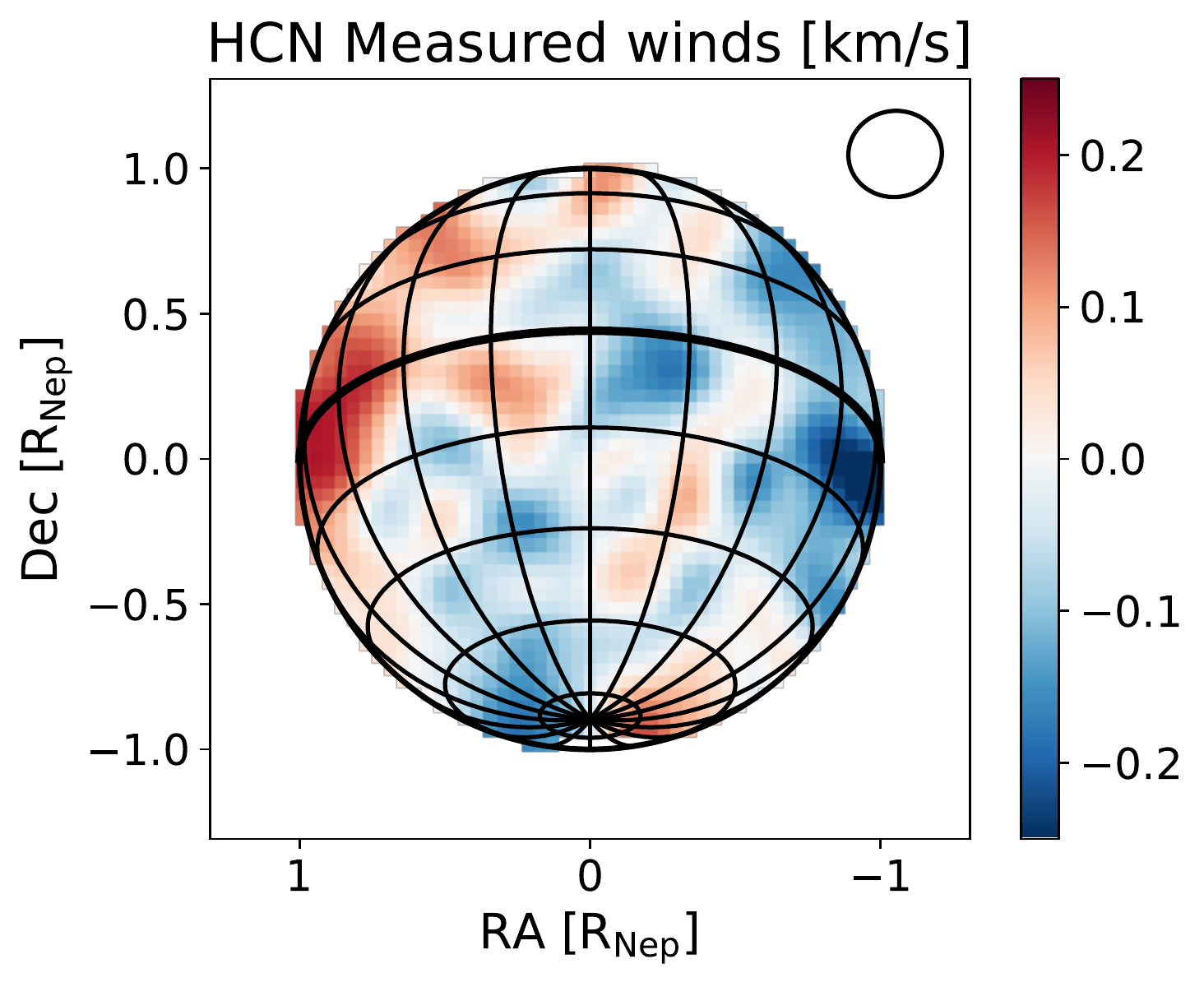} 
	\hfill
	\includegraphics[width=6.125cm]{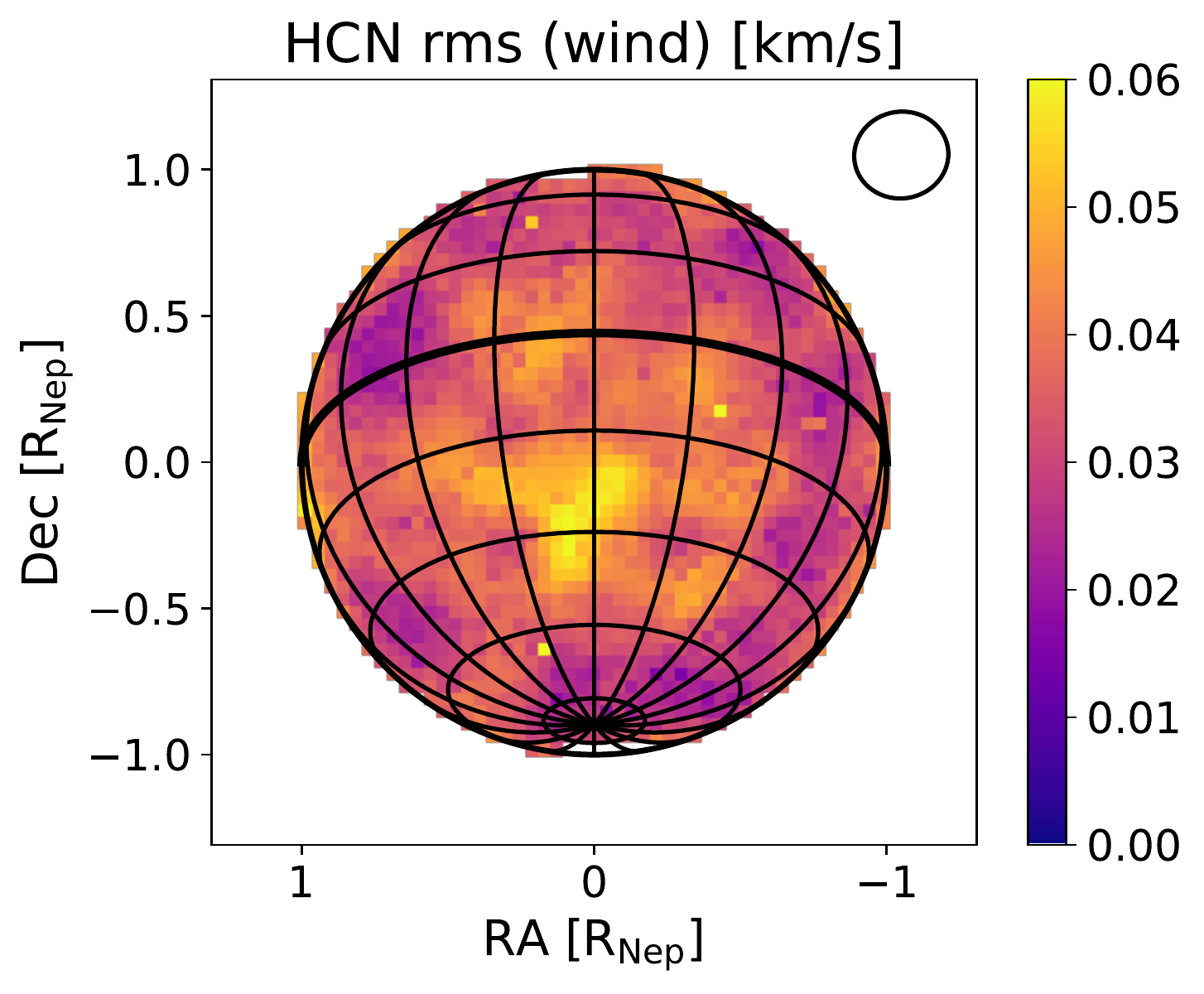} 
	  \\
	\caption{ALMA Doppler measurements for the CO(3-2) line (top row) and the HCN(4-3) line (bottom row). Left: measured Doppler lineshift. Middle: line-of-sight winds, after subtracting the solid-body rotation. Right: root-mean square (rms) of the measured winds. The top-right ellipse in each subplot shows the synthetic beam. Latitudes are shown in 20$^\circ$ steps, with the equator in a thicker line.}
\label{fig:maps_measurements}
\end{figure*} 

We observed Neptune on April 30, 2016, during 20 minutes on-source, with 41 antennas of the ALMA interferometer in the C36-2/3 hybrid configuration, yielding an angular resolution of about 0.37".
The spectral setup included the CO(3-2) line at 345.7959899 GHz and the HCN(4-3) line at 354.5054773 GHz, as well as the CS(7-6) line at 342.883 GHz \citep[whose detection was reported in][]{morenoetal2017}, with a spectral resolution of 1~MHz.

The bandpass, amplitude and phase calibration procedure  of the CS visibilities was described in \citet{morenoetal2017}, and we applied the same calibration procedure to the CO and HCN data using the ALMA/CASA data reduction software.

The resulting calibrated visibilities were then exported into
the GILDAS package to i) apply a self-calibration technique using Neptune’s continuum to improve image quality
ii) perform the imaging and  deconvolution using the H\"ogbom algorithm \citep{hogbom1974}.
We obtained a synthetic elliptical beam (robust weighting 0.5) of 0.39"$\times$0.34" (polar angle (PA) of -47$^{\circ}$) for CO and 0.37"$\times$0.35" (PA = -82$^{\circ}$) for HCN. The planet’s angular diameter was 2.24".
This yields a spatial resolution of about 20$^\circ$ at the equator.
The resulting spectral maps are shown Fig. \ref{fig:spectral_map_co_hcn} with signal-to-noise ratio (SNR) at line peak at the limbs of about 150 and 50, respectively for CO and HCN.
The final clean images were built with a sampling of 0.05" over -1.4" and +1.4" (3136 points).

An example of the observed lines is shown in Fig. \ref{fig:observations}.
The detailed analysis of the lineshapes in terms of spatial/vertical distribution of temperature, CO and HCN is left for future work. For the purpose of measuring winds from line central positions, we performed Gaussian fits: we used 3-Gaussian fits for CO (with initial FWHM of 4, 20, 80 km/s) and 2-Gaussian fits for HCN (with initial FWHM of 3, 16 km/s) as shown in Fig. \ref{fig:observations}.
We retained the narrowest component to measure the Doppler lineshift.
The high SNR in our maps allowed us to derive the Doppler lineshifts with an averaged velocity accuracy of 25 and 37 m/s, respectively for CO and HCN (Fig. \ref{fig:maps_measurements}, right column).

\section{Model} \label{sec:model}
\subsection{Radiative transfer} \label{subsec:model_RT}

We used the same Neptune radiative transfer model described in \citet{morenoetal2017}, as well as their thermal profile and their CO and HCN vertical distributions,  
to model the spectral lines of CO and HCN, 
and to compute the wind weighting function shown in Fig. \ref{fig:observations}.
Wind weighting functions account for the spectrally-dependent wind information content of each channel within lines \citep[see e.g.][]{lellouchetal2019}, and were convolved by the beam. 
At the limb, which is where most of the wind information comes from, CO lines probe the $2^{+12}_{-1.8}$~mbar level, and HCN probe the $0.4^{+0.5}_{-0.3}$~mbar level.

\subsection{Wind retrieval methodology} \label{subsec:model_retrieval}
In order to interpret the measured Doppler lineshifts, we developed a retrieval framework for the wind profiles based on the Markov-Chain Monte Carlo \texttt{emcee} sampler \citep{foremanmackeyetal2013}.
The observed lineshifts correspond to the sum of the line-of-sight Doppler displacement due to the planetary rotation and the winds. 
The planet rotation was modelled as solid-body rotation at the altitude of the 1~mbar level above the local planet radius.
Although somewhat different rotation periods have been proposed \citep{helledetal2010, karkoschka2011}, a period of 16.11~h \citep{warwicketal1989, lecacheuxetal1993} is adopted here to enable comparisons with the Voyager winds.

We parameterized the wind profiles ($W$), assumed purely zonal, as a polynomial function depending on the latitude ($\phi$, given in degrees) in the form:
\begin{equation}
W = \sum_n (a_n \times \phi^n)    
\end{equation}
and explored with \texttt{emcee} the parameter space described by the coefficients $a_n$, given in m/s.
The box priors which set the limits of the parameter space were [-500, 500] for the $a_0$ coefficient and [-1, 1] for the rest of coefficients. 
We tested polynomial orders from 0 to 5 for the wind parameterization, as discussed in Sect. \ref{sec:results}.

The MCMC sampler tests points of the $n$-dimensional parameter space of $a_n$ coefficients. 
For each test point, we computed the wind profile at latitudes $\phi \in [-90^\circ, 90^\circ]$ and the resulting wind map, with a pixel stepsize of 0.05". 
To simulate the line-of-sight Doppler lineshifts at each point of the map, we projected the zonal winds onto the planet geometry, accounting for the sub-observer latitude of 26.2ºS.
We then weighted the modelled lineshifts by the local intensity of the CO (resp. HCN) line and convolved by the ALMA beam, following a similar procedure to \citet{lellouchetal2019}. 
After adding the contribution of the solid body rotation, the test wind map ($W_{test}$) was compared with the one measured by ALMA ($W_{ALMA}$) for the molecule under study (Fig. \ref{fig:maps_measurements}, middle) by means of the $\chi^2$ figure of merit:

\begin{equation}
 \label{eq:chisq}
 \chi^2 =
\sum_{i=1}^{N_p}\left(\frac{W_{test}-W_{ALMA}}{rms_{ALMA}}  \right)^2  
\end{equation}
where $rms_{ALMA}$ is the 1-$\sigma$ error of the ALMA measurement for the molecule under study (Fig. \ref{fig:maps_measurements}, rightmost column) and $N_p$ is the number of pixels in the wind map.

We used 50 chains (or “walkers”) to simultaneously explore the parameter space independently in order to avoid possible $\chi^2$ local minima. The 50 walkers ran for up to 5$\times$10$^4$ steps, with a convergence criterion to stop the run at a number of steps larger than 50 times the autocorrelation time \citep[$\tau$, see ][for details]{foremanmackeyetal2013}. This convergence criterion ensures that the sampling of the parameter space has been completed and the sampled test points are effectively independent \citep{goodman-weare2010}. For the analysis of each retrieval run, we discarded the samples of the first $\tau$ steps (“burn-in phase”).

For each retrieval run, we defined the ensemble of good fits as those sampled wind profiles with a $\chi^2$ value in the 68.3\% (i.e. 1-$\sigma$) confidence level. That is, with $\chi^2$ verifying:
\begin{equation}
    \label{eq:confidence}
    \chi^2 - \chi^2_{min} \leq C \times \frac{\chi^2_{min}}{N} \times f_{oversampling}
\end{equation}
Here, $\chi^2_{min}$ is the minimum $\chi^2$ value among all the samples in the run, and $\chi^2_{min}/N$ is the reduced value of $\chi^2_{min}$, with $N$ equal to the degrees of freedom of the retrieval ($N=N_p-n$).
The factor $f_{oversampling}=Area_{beam}/Area_{pixel}$ accounts for the fact that with a spatial sampling step of 0.05" and a beam of 0.37", the measurements are considerably oversampled, and the number of independent measurements is $N_p/f_{oversampling}$.
The coefficient $C$ denotes the 1-$\sigma$ confidence region in the $n$-dimensional parameter space, where $n$ is the number of $a_n$ coefficients in the polynomial fit.
For polynomial orders 0 to 5, $C$ is 1.0, 2.3, 3.53, 4.72, 5.89, and 7.04, respectively \citep[see Ch. 15.6 in][]{pressetal2007}.

\section{Results} \label{sec:results}

\begin{table*}
\centering
\caption{Polynomial coefficients of the best fits to the CO and HCN zonal wind measurements. Also given, the coefficients that approximate the envelope of good fits to a fourth-order polynomial.}
\label{table:results_coefficients} 
\begin{tabular}{l c c c c c}
\hline 
\hline
    &    $a_0$   & $a_1$   & $a_2$   & $a_3$   & $a_4$    \\
\hline
CO best fit & -1.81$\times10^{+2}$ & 8.76$\times10^{-1}$ & 6.04$\times10^{-2}$ & -7.44$\times10^{-4}$ & -5.50$\times10^{-6}$ \\
CO upper error & -1.29$\times10^{+2}$ & -3.25$\times10^{-1}$ & 5.42$\times10^{-2}$ & 1.41$\times10^{-3}$ & 2.72$\times10^{-5}$ \\
CO lower error & -2.28$\times10^{+2}$ & 8.53$\times10^{-1}$ & 1.42$\times10^{-2}$ & -2.89$\times10^{-3}$ & -3.06$\times10^{-5}$ \\
HCN best fit & -1.87$\times10^{+2}$ & -8.11$\times10^{-1}$ & 8.39$\times10^{-2}$ & 4.53$\times10^{-4}$ & 2.14$\times10^{-6}$ \\
HCN upper error & -1.07$\times10^{+2}$ & -7.46$\times10^{-1}$ & 6.61$\times10^{-2}$ & 1.29$\times10^{-3}$ & 2.11$\times10^{-5}$ \\
HCN lower error & -2.40$\times10^{+2}$ & 4.92$\times10^{-1}$ & 7.47$\times10^{-2}$ & -1.79$\times10^{-3}$ & -2.92$\times10^{-5}$ \\
\hline
\end{tabular}
\end{table*}

\begin{figure*}
	\centering
    \includegraphics[width=4.85cm]{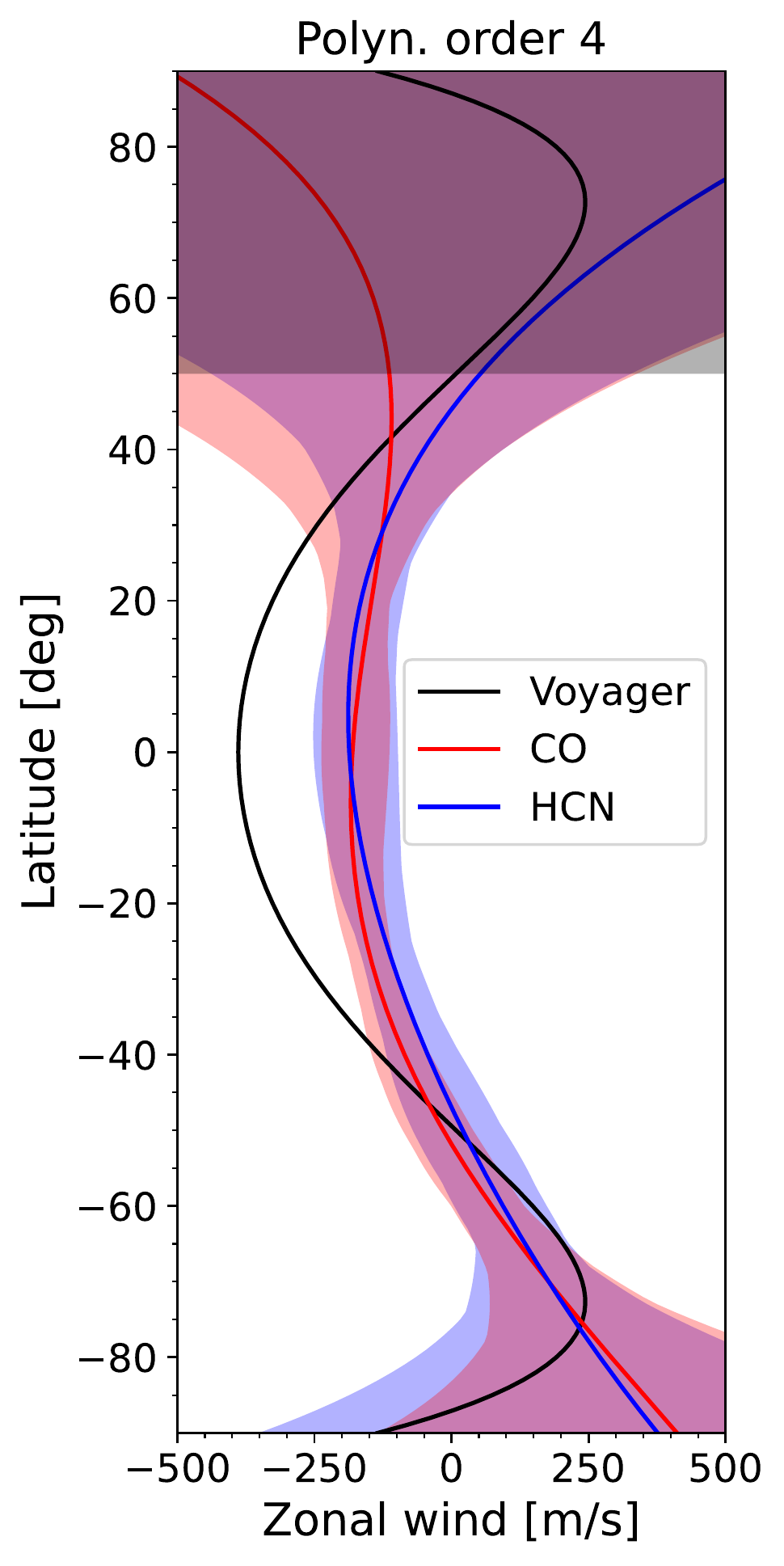} 
    \includegraphics[width=13.15cm]{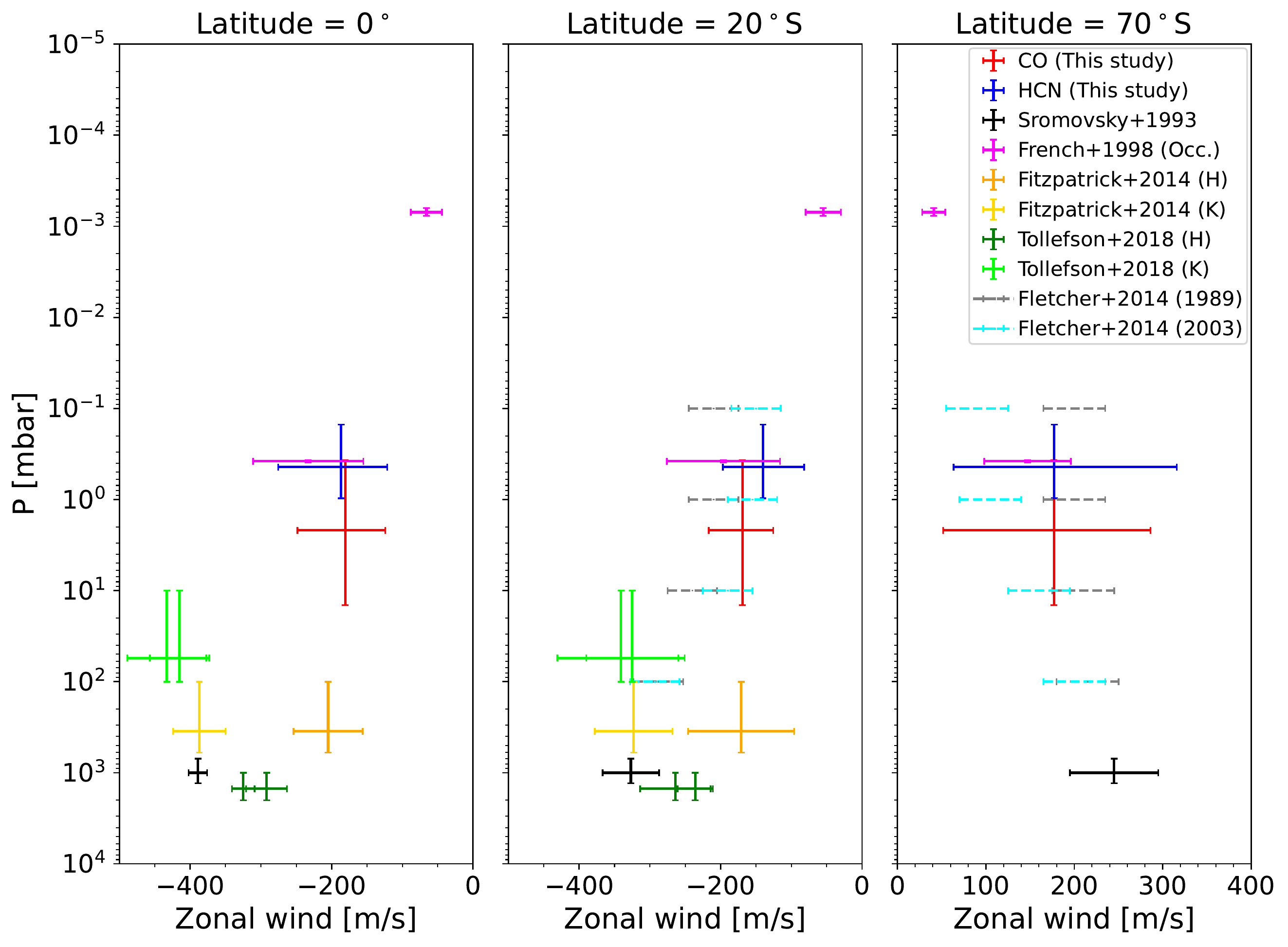}
	\caption{Retrieval results. Left: Retrieved best-fitting wind profiles for CO (red line) and HCN (blue line) measurements, using a fourth-order polynomial. Red and blue shadowed regions contain the ensemble of good fits according to the $\chi^2$ criterion from Eq. (\ref{eq:confidence}). The semi-transparent grey rectangle indicates the unobservable northern latitudes. The solid black line shows the sixth-order fit to Voyager's cloud-tracking winds \citep{sromovskyetal1993}. Right: Wind variations with altitude at the equator, 20ºS and 70ºS, both for our measurements and for a set of references in the literature (see Sect. \ref{sec:discussion} for details). Cloud-tracking winds from \citet{fitzpatricketal2014} and \citet{tollefsonetal2018} are not shown at 70$^\circ$S as they have uncertainties of about 1000~m/s. Winds from \citet{fletcheretal2014} do not correspond to a direct measurement, but to the computed thermal wind equation applied to a reanalysis of the 1989 IRIS/Voyager data (dashed grey lines) and to 2003 Keck data (dashed cyan lines).}
\label{fig:results_reference}
\end{figure*} 

With the method above, we carried out wind retrievals for the CO and HCN measurements.
We ran retrievals for wind parameterizations with polynomials of orders between 0 and 5.
Previous cloud-tracking studies had generally assumed latitudinally symmetric wind profiles and thus omitted odd polynomial orders \citep{sromovskyetal1993, fitzpatricketal2014, tollefsonetal2018}.
However, the spatial resolution achieved in our data is potentially sensitive to latitudinal wind asymmetries.
We therefore kept odd polynomial orders in our retrievals.

The retrieval results for CO and HCN with polynomial fits of orders 0 to 5 are shown in Fig. \ref{fig:appendix_windprofiles}.
The 6th-order polynomial fit to Voyager's cloud-tracking measurements \citep{sromovskyetal1993} is shown for comparison.
For each tested polynomial order, we include in Fig. \ref{fig:appendix_windprofiles} the map of residuals between the observed Doppler lineshifts and the best-fitting model (solid rotation plus wind) to assess the variation of fit quality as a function of polynomial degree.
For both CO and HCN, we find that the value of $\chi^2/N$ is practically the same for retrievals or orders 3, 4 and 5.
This implies that these polynomials are similarly able to fit the measurements at the latitudes we are sensitive to.
Indeed, the three parameterizations retrieve similar values of the zonal winds at latitudes over 20$^\circ$N to 70$^\circ$S.

Given Neptune's sub-observer latitude of 26.2°S and projection effects, wind information is restricted to latitudes southward of ~40°N.
In addition, the beam size of 0.37" prevents detailed information at southern polar latitudes. Angular speed considerations indicate that the wind speed should theoretically be zero at the pole. We find that this constraint is met within error bars for 4th and 5th-order polynomial fits.
Order $n$=4 thus represents a good compromise between the model's mathematical complexity and physical realism, and we adopt it as the reference wind parameterization.

Figure \ref{fig:results_reference} shows the retrieved wind profiles for the reference $n$=4 parameterization.
Table \ref{table:results_coefficients} shows the polynomial coefficients for CO and HCN best-fit wind profile, as well as the coefficients that parameterize the envelope of good solutions --those verifying Eq.~(\ref{eq:confidence})-- using a fourth-order polynomial fit.
Fig. \ref{fig:results_reference} also shows our retrieved wind measurements at 0$^\circ$, 20$^\circ$ and 70$^\circ$S as a function of pressure.
For reference, the plot includes previously reported zonal winds (see discussion in Sect. \ref{sec:discussion}).

At the equator, we retrieve retrograde (westward) zonal winds of $-180 ^{+70}_{-60}$~m/s from CO measurements, and $-190 ^{+90}_{-70}$~m/s from HCN.
We find a decrease in the wind intensity towards mid-latitudes.
At 20$^\circ$S, wind speeds are $-170 ^{+50}_{-40} $~m/s for CO ($-140 ^{+60}_{-60} $~m/s for HCN), and at 40$^\circ$S they are $-90 ^{+50}_{-60} $~m/s for CO ($-40 ^{+60}_{-80} $~m/s for HCN).
Winds continue this trend towards southern latitudes, becoming 0~m/s at about 50$^\circ$S.
At 70$^\circ$S, winds are prograde (eastward) although uncertainties are larger ($180 ^{+130}_{-110} $~m/s for CO, and $180 ^{+110}_{-140} $~m/s for HCN).
Further South than 70$^\circ$S, winds remain unconstrained due to the limited spatial resolution.
Wind speeds in the observable northern-hemisphere latitudes are compatible, within errors, with a symmetric wind profile.
At 20$^\circ$N, zonal winds are $-150 ^{+30}_{-80} $~m/s for CO and $-170 ^{+80}_{-50} $~m/s for HCN.
Additional measurements will be needed to confirm this behaviour, eventually when regions further North become observable.

\section{Discussion}    \label{sec:discussion}

Our results provide a new method to probe Neptune's stratospheric winds and wind shear, also yielding latitudinal information. 
In Fig. \ref{fig:results_reference}, representative values of wind speeds at 0°, 20°, and 70°S from CO and HCN are plotted in the context of previous measurements. 
Within errors, our values agree with thermal winds calculated by \citet{fletcheretal2014}, both for their reanalysis of the Voyager/IRIS data and for their 2003 Keck data.
Our retrieved winds also agree with the wind speed at 0.38 mbar derived by \citet{frenchetal1998} from stellar occultations. 
Comparison with the Voyager 2 cloud-tracking winds \citep{sromovskyetal1993} also confirms a decay of the wind intensity with latitude, and a smaller wind shear at high latitudes. 
Although the various observations pertain to different epochs, our results validate predictions of the thermal wind equation and argue for a preservation of the general circulation pattern over the 28 year interval ($\sim$60 degrees in heliocentric longitude) spanned by the data around the 2005 southern summer solstice.

Our equatorial wind is about $200^{+100}_{-80}$~m/s less intense than the Voyager reference.
Assuming nominal probed levels of $\sim$1~mbar and $\sim$1~bar respectively, this indicates a $+70^{+30}_{-20}$~m/s wind shear per pressure decade, (or $\partial u / \partial z = +30 \pm 10$m/s per scale height) where the positive sign is related to the retrograde wind direction. 
At 70ºS, our winds are about $70^{+180}_{-170}$~m/s less intense than Voyager's. We find a much smaller wind shear at 70ºS, although the uncertainties are larger than for equatorial winds: $-20\pm60$~m/s per pressure decade (or $\partial u / \partial z = -9 \pm 25$m/s per scale height).  
Our results compare well with the estimates from \citet[their Fig. 11b]{frenchetal1998}, who studied the wind-shear between Voyager's cloud-tracking winds (which they assumed at 100~mbar) and  their occultation data at 0.38~mbar.
\citet{frenchetal1998} determined a wind shear of about +30~m/s per scale height at the equator, and -15~m/s per scale height at 70ºS.

In contrast, cloud-tracking measurements from \citet{tollefsonetal2018} appear somewhat at odds with our estimated wind shear, as their H band measurements --assumed by the authors to probe deeper levels-- indicate less intense winds than the K' band.
\citet{tollefsonetal2018} assumed that the H-band (resp. K' winds) probe the 1-2 bar (resp. 10-100 mbar) level.
This led them to an inverted wind shear, with equatorial winds becoming more intense with increasing altitude.
They attempted to explain this behavior by invoking a thermal-compositional wind equation that accounts for density changes associated to latitudinal variations of the methane abundance. Such an approach is not warranted according to our results. 
Furthermore, the absolute sounded pressures are uncertain and highly model-dependent, and both bands might not be probing such different pressure levels \citep[Fig. 16 therein]{tollefsonetal2018}.  
Similarly, \citet{fitzpatricketal2014} find differences between their cloud-tracking H and K'-band winds.
The pressure levels of the observed clouds are also uncertain in this case, with both H and K'-band clouds spanning pressure levels between 0.1-0.6~bar \citep[Fig. 11 therein]{fitzpatricketal2014}.

In itself, the consistency of our direct wind measurements with thermal wind calculations does not highlight a particular mechanism responsible for the wind decay with altitude: the thermal wind equation simply states a balance between vertical wind shears and temperature meridional gradients. The wind decay reported here indicates a drag source, which could be the propagation and breaking of gravity and/or planetary waves  (common in planetary stratospheres), although this has to be tested in dynamical simulations. On Saturn and Jupiter, interactions between vertically-propagating waves and the mean zonal flow drive strong acceleration and deceleration of the stratospheric equatorial zonal flow \citep[e.g.][]{cosentinoetal2017, bardetetal2022}. Wave breaking as a source of friction was also hypothesized by \citet{ingersolletal2021} to maintain the stacked circulation cells in Jupiter’s upper troposphere.

Our measurements open a new window in the study of Neptune's stratospheric dynamics.
Also, our findings provide useful information for general-circulation modelling studies \citep{liu-schneider2010, milcarecketal2021}, which require observations to compare with the outputs of the numerical simulations.
Our wind measurements remain nevertheless modest in precision, as a result of combined limited integration time and low spatial resolution. 
Future dedicated observations, possibly combined with long-term monitoring (given the duration of Neptune seasons), are expected to yield further insight into the topic.

\begin{acknowledgements}

This paper makes use of the following ALMA data:
ADS/JAO.ALMA\#2015.1.01471.S. (PI: R. Moreno). ALMA is a partnership of ESO (representing its member states),
NSF (USA) and NINS (Japan), together with NRC (Canada), NSC and ASIAA (Taiwan), and KASI
(Republic of Korea), in cooperation with the Republic of Chile.
The Joint ALMA Observatory is operated by ESO, AUI/NRAO and NAOJ. 
The authors acknowledge the support of the French Agence Nationale de la Recherche (ANR), under grant ANR-20-CE49-0009 (project SOUND).
T. Cavali\'e acknowledges funding from CNES.
\end{acknowledgements}

%
%

\bibliographystyle{aa}
\bibliography{references}
\clearpage

\begin{appendix} 

\begin{figure*}
\section{Measured spectral maps} \label{appendix:sec_spectral_maps}
\centering
    \includegraphics[width=18cm]{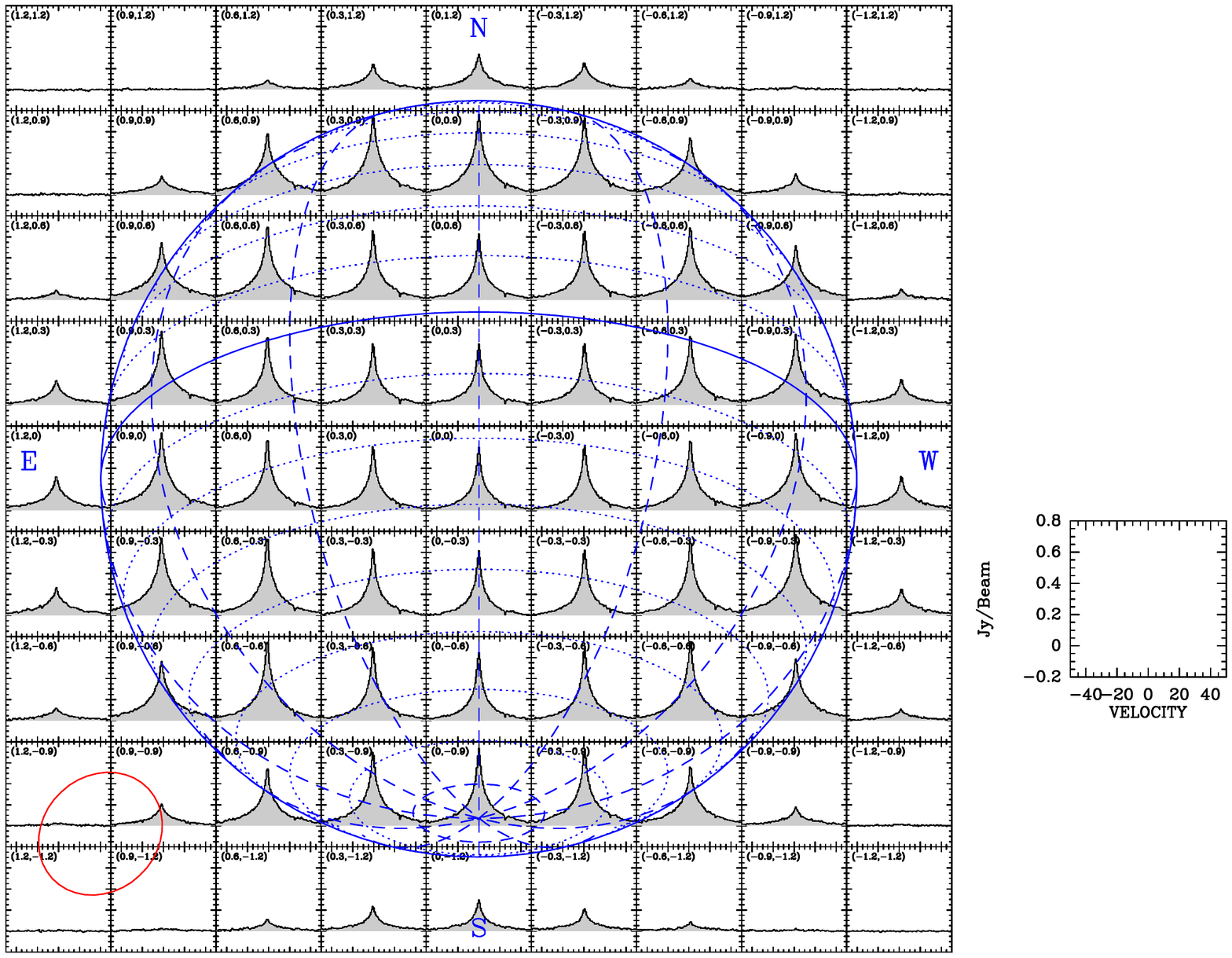}
    \vspace{-1.5cm}
    \\
    \includegraphics[width=18cm]{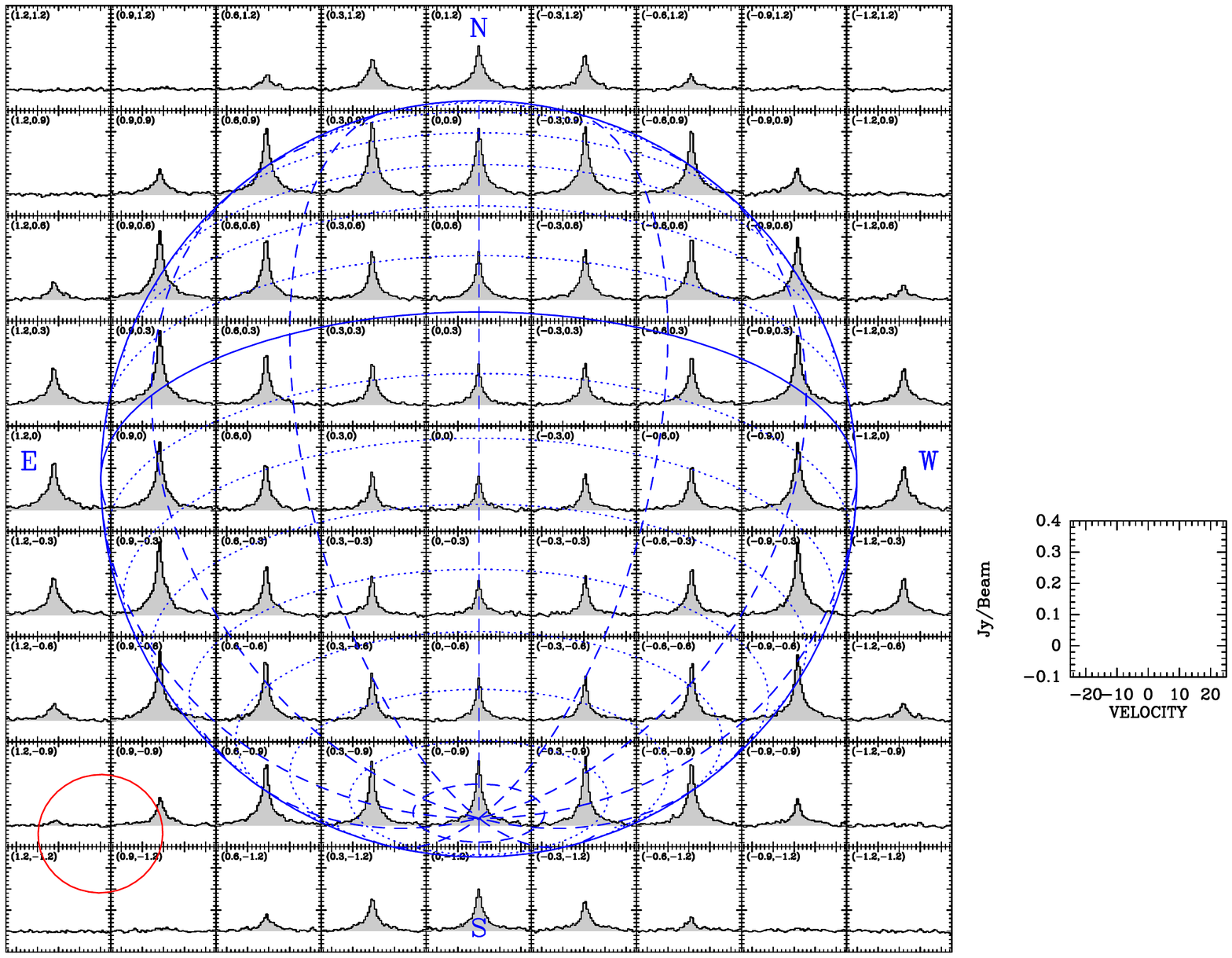}
    \vspace{-1.5cm}
	\caption{Measured spectral maps. Top: Spectral map of the CO(3--2) transition at 345.796 GHz on Neptune observed on April 30, 2016 with ALMA (top row). The spectral resolution is 1 MHz. The planet’s angular diameter is 2.24" and is shown by a blue circle. The synthesized beam is indicated with a red ellipse. The right box represent the scales of the spectra expressed in flux (Jy/beam) against velocity (km/s). Bottom: Spectral map of the HCN(4--3) transition at 354.505 GHz, measured simultaneously with that of CO.}
\label{fig:spectral_map_co_hcn}
\end{figure*}
\FloatBarrier

\begin{figure*}
\section{Parameterizing wind profiles with different polynomial orders} \label{appendix:sec:retrievals_polyn-orders}
	\centering
    \includegraphics[width=18cm]{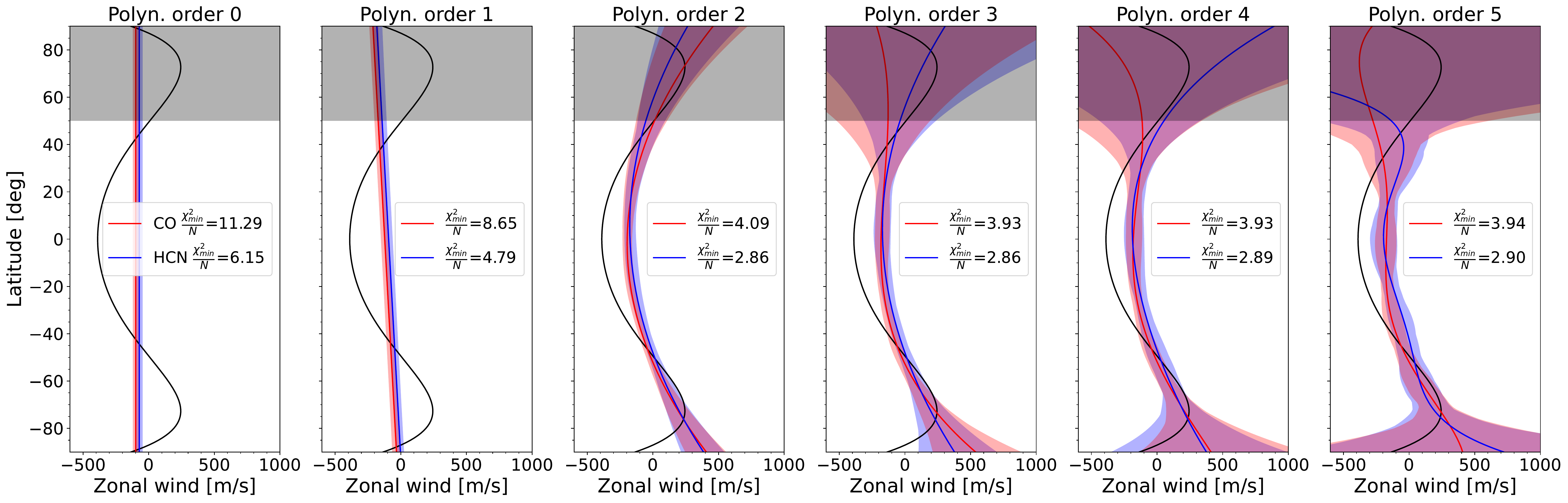}
    \\
    \includegraphics[width=18cm]{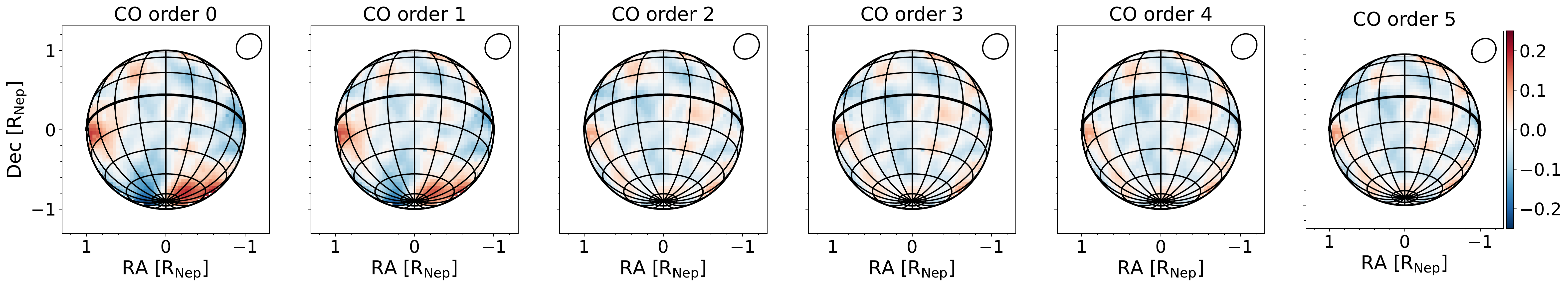}
    \\
    \includegraphics[width=18cm]{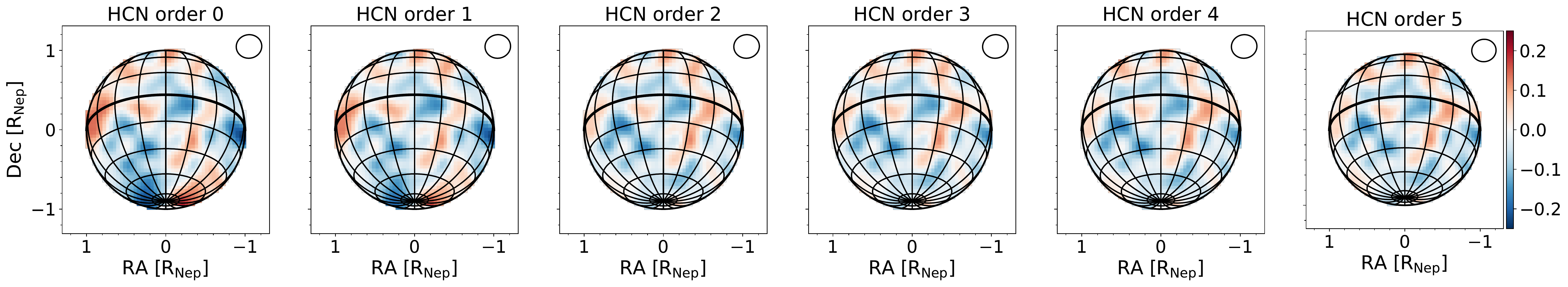}
	\caption{As Fig. \ref{fig:results_reference}, the top row shows the retrieved wind profiles for CO and HCN but using different polynomial orders for the wind parameterization. The value of $\chi^2_{min}/N$ is shown in the legend. We also show the residual maps of CO (middle row) and HCN (bottom row) after subtracting the best-fitting wind profile to the measurements.}
\label{fig:appendix_windprofiles}
\end{figure*} 
\FloatBarrier

\end{appendix}

\end{document}